\documentclass[fleqn,usenatbib]{mnras}

\usepackage{newtxtext,newtxmath}

\usepackage[T1]{fontenc}

\DeclareRobustCommand{\VAN}[3]{#2}
\let\VANthebibliography\thebibliography
\def\thebibliography{\DeclareRobustCommand{\VAN}[3]{##3}\VANthebibliography}

\usepackage{graphicx}	
\usepackage{amsmath}	

\defcitealias{Krumholz2012}{KDM12}

\title[Forming Stars in Mrk~266]{Forming Stars in a Dual AGN Host: Molecular and Ionized Gas in the Nearby, Luminous Infrared Merger, Mrk~266}

\author[D. Beaulieu et al.]{Damien Beaulieu$^{1,2}$\thanks{E-mail: damien.beaulieu.1@ulaval.ca},
Andreea Petric$^{3}$\thanks{E-mail: apetric@stsci.edu},
Carmelle Robert$^{1,2}$,
Katherine Alatalo${^3}$,
Timothy Heckman$^{4}$,
\newauthor{Maya Merhi}$^{5}$,
Laurie Rousseau-Nepton${^6}$,
Kate Rowlands$^{3,4}$
\\
$^1$ Département de physique, de génie physique et d'optique, Université Laval, 1045 Av. de la Médecine, Québec, G1V 06A, Canada\\
$^{2}$ Centre de Recherche en Astrophysique du Québec\\
$^{3}$ Space Telescope Science Institute, 3700 San Martin Drive, Baltimore, MD 21211, USA\\
$^{4}$ Department of Physics and Astronomy, Johns Hopkins University, Baltimore, MD 21218 USA\\
$^{5}$ Center for Social Data Analytics, Pennsylvania State University, 201 Old Main, State College, PA 16802, USA\\
$^{6}$ Canada-France-Hawaii Telescope, 65-1238 Mamalahoa Highway, Kamuela, HI 96743, USA\\
}

\date{Accepted XXX. Received YYY; in original form ZZZ}

\pubyear{2022}

\begin{document}
\label{firstpage}
\pagerange{\pageref{firstpage}--\pageref{lastpage}}
\maketitle

\begin{abstract}
We present star formation rates based on cold and ionized gas measurements of Mrk~266 (NGC\,5256), a system composed of two colliding gas-rich galaxies, each hosting an active galactic nucleus. Using $^{12}$CO\,(1-0) observations with the Combined Array for Research in Millimeter-Wave Astronomy (CARMA), we find a total H$_2$ mass in the central region of $1.1\pm0.3\times10^{10}$~$M_\odot$ which leads to a possible future star formation rate of $25\pm10~M_\odot$~yr$^{-1}$. With the Fourier Transform Spectrograph (SITELLE) on the Canada-France-Hawaii Telescope, we measure an integrated H$\alpha$ luminosity and estimate a present-day star formation rate of $15\pm2~M_\odot$~yr$^{-1}$ in the core of the system (avoiding the two active nuclei). These results confirm that Mrk~266 is an intermediate stage merger with a relatively high recent star formation rate  and enough molecular gas to sustain it for a few hundred million years. Inflowing gas associated with the merger may have triggered both the starburst episode and two AGN but the two galaxy components differ: the region around the SW nucleus appears to be more active than the NE nucleus, which seems relatively quiet. We speculate that this difference may originate in the properties of the interstellar medium in the two systems.
\end{abstract}

\begin{keywords}
Galaxies:individual: NGC\,5256 -- galaxies: star formation -- galaxies: nuclei -- galaxies: interactions 
\end{keywords}



\section{Introduction} \label{introduction}
Interactions play a significant role in the evolution of galaxies. The changing gravitational potential produces tidal shocks and flows that compress and heat the gas, enhancing star formation and disturbing its orbital motion \citep{Kennicutt1987,Mihos1996,sanders1996, Veilleux2006, Petric2011, Ellison2013a, Madau2014, Rich2014, morganti2017, ala2019}. The associated gravitational torque removes angular momentum from the gas, allowing some of it to be accreted onto supermassive black holes (SMBHs), triggering activity observed as Active Galactic Nuclei (AGN).
The resulting feedback from the AGN associated phenomena (e.g., radiation, jets and winds) and star formation may heat the dust and produce relatively higher IR luminosities \cite[]{sanders1996,pt2021}.\\

Luminous Infrared Galaxies (LIRGs) are defined as galaxies with $L_\text{IR} \geq 10^{11} L_\odot$ \citep{sanders1996}. The Great Observatories All-sky LIRG Survey (GOALS: \citealt{armus2009}) comprises a complete, volume-limited sample of nearby LIRGs and ULIRGs from isolated disks to major merger remnants. The GOALS sample is ideal for studying the evolution of galaxies because of multiwavelength observations allowing the detection of dust, ionized gas, molecular gas, and young and old stellar populations. Of the 202 objects in GOALS, half are interacting systems, from minor to major mergers, and 18\% host AGN  \cite[]{larson2016,stierwalt2014,inami2013,Petric2011}. AGN feedback can affect star formation in the host galaxy via shocks and turbulence, which increase the temperature structure of the surrounding ISM \citep[]{Alatalo2015, morganti2017, Petric2018, lamb2019, minsley2020, riff2020}, therefore LIRGs are ideal for studying the impact of AGN on star formation and the interstellar medium (ISM). AGNs accrete large quantities of gas, but a significant proportion of the molecular gas, the raw material of star formation, can be pushed out of the galaxy. The galaxy gravitational potential may also bring back the ejected gas, enhancing the star formation rate \cite[SFR,][]{Alatalo2015}. The observed, tight correlation between the SMBH mass and the host galaxy properties suggests that AGN and star-formation processes both impact the ISM and thus each others growth \citep[]{Kormedy2013, Heckman2014}.\\

Molecular and atomic hydrogen indicate the amount of gas available for star formation. Therefore, we can use the global properties of the cold gas to estimate the future star formation rate if we make assumptions about the efficiency of the process \cite[e.g.][]{Krumholz2012}. Furthermore, the amount of observed molecular (and atomic) hydrogen correlates with the observed present-day star formation rate measured from either the H$\alpha$ or mid-IR luminosity. This relation is known as the Schmidt-Kennicutt law \citep{Schmidt1959, Kennicutt1998}. The present-day star formation rate indicators originate in regions of young stellar populations where massive stars ionize and heat the surrounding gas and dust clouds. \citet{Kennicutt2021}  use a sample of galaxies ranging from non-starbursting to starburst galaxies to refine the slope of the relation between the star-formation surface density $\Sigma_{\text{SFR(IR)}}$ and the molecular gas surface density $\Sigma_{\text{H}_2}$. They 
find different power laws for the two galaxy groups (normal and starbursts), suggesting that the Schmidt-Kennicutt law is bimodal or even multimodal. \cite{Kennicutt2021} propose that this behavior could originate in a change of the small-scale structure of the ISM: a mix of atomic and molecular hydrogen in normal disk galaxies as opposed to a molecular-dominated ISM in starburst galaxies, which supports the conclusions of \citet{Bigiel2008, Bigiel2011} and \cite{Leroy2008}.\\

High resolution and sensitive observations of molecular gas in AGN hosts show that gravitational torque, most often associated with a nuclear bar that supports the angular momentum exchange, can increase accretion onto the SMBH. Kiloparscec and parsec scale observations point to coupled kinematics between AGN feedback, star formation and the host galaxy ISM through inflows in a thin disk around the AGN and outflows in the direction perpendicular to the disk \citep{Alatalo2015, ala2019, aud2019, combes2019}. During the merging process, inflowing gas may grow both SMBHs, making the system the host of a dual AGN \citep{Hopkins2006,Steinborn2016,Solanes2019}. The dual AGN stage happens when the two nuclei are close companions, i.e., during the late stages of a merger when accretion rates are at their peak \citep{Ellison2013b,Capelo2017,Weigel2018,Koss2019,Solanes2019}. Most, if not all, major mergers should, at one point in their evolution from two separate galaxies to a single one, host a dual AGN \cite[]{Koss2019} however, only a few know dual AGN systems have been observed \cite[]{Petric2011}.\\

The most distant, luminous, and dustiest quasars show the marks of gravitational interactions from tidal tails to complex nuclear structures. \cite{urrutia2008}, \cite{glikman2015} and \cite{lacy2018} find that the hosts of obscured AGNs tend to be major mergers. The observations of \citet{glikman2012,glikman2015} are particularly telling: at the peak epoch for galaxy, and BH growth, the most luminous quasars are also the most dust reddened and major mergers. These compelling findings pave the way for the next decade of discoveries. The next generation of X-ray, radio, and IR wide field/all-sky surveys will target obscured quasars to get a statistically sound handle on their demographics, star formation rates, and stellar population histories. Imaging surveys will require follow-up with efficient (i.e., sensitive, wide aperture, and highly multiplexed) spectroscopic surveys in the optical and NIR \citep{petric2019}. Such facilities will allow us to measure the integrated properties of obscured AGN in merging systems like Mrk~266. Studying local counterparts to this high redshift population of obscured AGNs is fundamental for understanding the relevant physical processes at the smallest scales.\\

\begin{figure*}
    \centering
    \includegraphics[width = 0.80\textwidth]{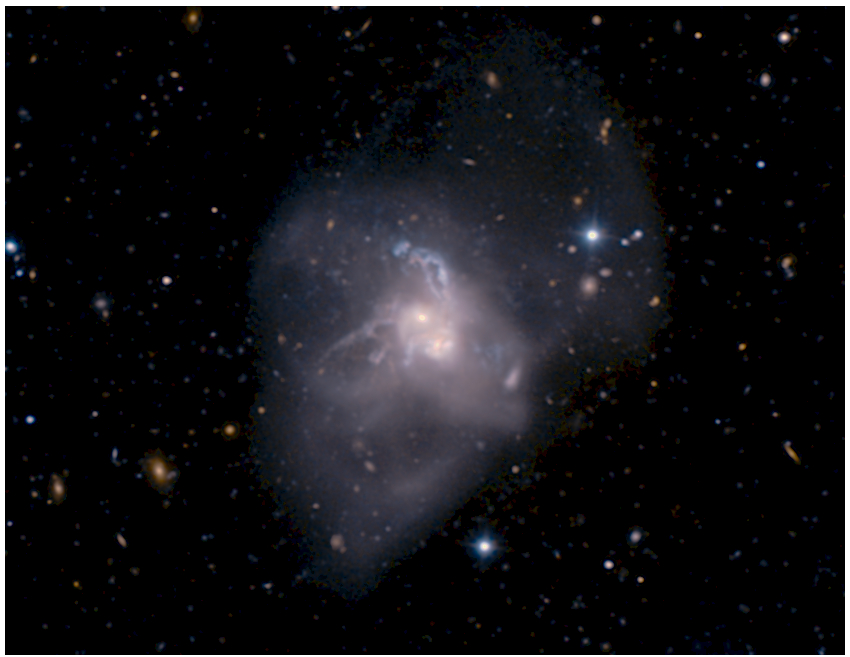}
    \caption{ Deep image of Mrk~266 as observed with SITELLE. Each pixel is colored by combining data from the SN2 filter (482-513~nm) in blue and the SN3 filter (647-685~nm) in red. This final image is cropped to a size of 4.5'x3.5'. East is left and North is up.
    \label{image_Mrk266}}
\end{figure*}

Mrk~266 (NGC~5256) has attracted considerable attention due to its particular optical morphology. \cite{Zwicky1971} first noticed a double-nucleus structure and \citep{Osterbrock1983} classified the south-western (SW) nucleus as a Seyfert 2 and the nucleus at the north-east (NE) as a LINER. The nuclei are separated by 10" which at the estimated distance of 129~Mpc (z=0.0279) is 6~kpc \cite[]{Iwasawa2020}. Figure~\ref{image_Mrk266}, an optical deep image of Mrk~266 displays signs of a disturbed morphology; the two nuclei are distinguishable and structures such as plumes \cite{hut1988}, tidal tails \cite[]{Brassington2007}, knots \citep{Mazzarella1988,hut1988,Ishigaki2000, Mazzarella2012}, winds \citep{Wang1997, Kollatschny1998, Ishigaki2000, Davies2000, Brassington2007,Mazzarella2012} and an extended but faint ionized hydrogen envelope \citep{Armus1990,Wang1997}. These signs all indicate that Mrk~266 is an ongoing major merger between two galaxies of similar masses. In addition, the galaxy appears to be in a "blow-out phase" \cite[]{Mazzarella2012} where galactic-scale superwind is sweeping massive amount of dust revealing previously obscured features \cite{hut1988,Ishigaki2000}. Based on these observations, merger simulations \citep{Mihos1996, Hopkins2006} and nucleus separation, the final coalescence of the SMBH will take place in 50~to~300~Myr  \citep{Brassington2007, Mazzarella2012}. For these reasons, the galaxy is considered to be in an intermediate merging stage. Furthermore, Mrk~266 presents an excess of IR with a total FIR luminosity of $L_{8-1000\mu\text{m}}=10^{11.56}L_\odot$ \cite[]{armus2009}. It has already been established that there is a clear connection between mergers, (U)LIRG and AGN \citep{Sanders1988, Mihos1996, sanders1996, Petric2011}. The SW region dominates the MIR and FIR emission of the system \cite[]{Mazzarella2012} but a contribution from the NE region may be non-negligible \cite[]{Ishigaki2000}, however, the origin of the IR luminosity of Mrk~266 is quite complex as dust is heated by a combination of star formation and of AGN activity \citep{Mizutani1994, Hwang1999, Ishigaki2000, Brassington2007, Mazzarella2012}. \\

More surprisingly, both the SW and the NE nucleus manifest AGN signatures. The dual AGN nature of Mrk~266 has been suggested by many studies \citep{Wang1997, Mazzarella1988, Davies2000, Imanishi2009, Wang2010}, but it had not been confirmed until the multiwavelength investigation of \cite{Mazzarella2012}. Despite having similar SMBH masses of $\sim2\times10^{8} M_\odot$ \citep[]{Marconi2003, Mazzarella2012} and being in close proximity, the nucleus properties are considerably different and may be caused by unbalanced accretion conditions \cite[]{Iwasawa2020}. Mrk~266 SW is 50\% more luminous than the average galaxy of the local universe and based on Hubble Space Telescope (HST) observations, its progenitor is a gas-rich barred spiral (SBb) with two discernable spiral arms \cite[]{Mazzarella2012}. Hard X-ray observations have confirmed that the SW nucleus is a heavily obscured Compton-thick AGN and is intrinsically 30~times more luminous than the NE nucleus \cite[]{Iwasawa2020}. Polycyclic aromatic hydrocarbons (PAH) are detected in the SW region \citep{Davies2000, Mazzarella2012, Iwasawa2020} and young to intermediate stars dominate the  continuum emission, \cite{Wang1997} indicate that active star formation is taking place in this region. On the other hand, Mrk~266 NE is 30\% more luminous than the average galaxy of the local universe and its progenitor is thought to be a relatively gas-poor S0/a; on its own (i.e. pre-merger) the NE galaxy would not have sustained high star formation rates \cite[]{Mazzarella2012}. Because the NE nucleus is brighter than the SW nucleus in X-ray, UV and optical, it is considered to be less obscured than the SW but it is still a moderately obscured low-luminosity AGN \cite{Iwasawa2020}. Indicators like PAH \citep{Davies2000, Mazzarella2012, Iwasawa2020} and hot plasma \cite[]{Wang2010} are observed which suggest a circumnuclear starburst in the NE region. Overall, the spectral energy distribution (SED) analysis of \cite{Mazzarella2012} suggests that Mrk~266 is starburst dominated, but the bolometric luminosity is equally powered by AGN and starburst \citep{Wang1997,Mazzarella2012}. \\

The region between the nuclei is also of great interest. Located halfway between the two nuclei, there is a curious radio source with no optical counterpart \citep{hut1988, Mazzarella1988, Kollatschny1998, Davies2000, Mazzarella2012}. Although many explanations are possible, previous studies argue that it is an interface region between the two colliding galaxies where shock-induced synchrotron emission is produced \cite[]{Mazzarella2012}. Another interesting phenomenon observed in the central region of Mrk~266 and perhaps the main focus of this study is a molecular gas bridge that connects the two nuclei \cite[]{Imanishi2009}. The presence of such a massive gas reservoir is an indication that the central region of Mrk~266 may host a powerful starburst that is to be sustained. It is also an indication that molecular gas is being transferred from one nucleus to the other \cite[]{Cox2006}. The observation of molecular gas bridges remain limited since only three other galaxies of GOALS (VV~114, NGC~6090 and NGC~6240) have a significant amount of molecular gas between their nuclei \cite{Mazzarella2012}. \\

Mrk~266 is an ideal laboratory to study the connection between dual AGN and star formation because it is in a merging phase where two AGN and a powerful starburst are all sustained and witnessing these two phenomena simultaneously is a rare occurrence because they are both short time scale events. This is possibly an important stage in the merging process where LIRGs turns into ULIRGs and investigating the impact of a dual AGN on its host ability to form new stars is key to better comprehend how galaxies evolve through mergers.\\

In this paper, we present a study the molecular and ionized gas emission in the central region of the two merging galaxies in Mrk~266 and we mainly focus on the impact of a dual AGN on the host ability to form new stars. CO and H$\alpha$ observations used to compute SFR are presented in Section~\ref{observations}. CO velocities are used to compute dynamical masses in Section~\ref{COtodyn} while CO fluxes are used to compute H$_2$ masses in Section~\ref{COtomass}. Various methods are used to compute SFRs; Section~\ref{surftoSFR} presents methods utilize H$_2$ surface density and Section~\ref{SFRfromHa} presents methods based on H$\alpha$ emission. We discuss the implications of the computed SFRs in Section~\ref{discussion} and conclude in Section~\ref{conclusion}.


\section{Observations} \label{observations}

\begin{figure*}
    \centering
    \includegraphics[width = 0.9\textwidth]{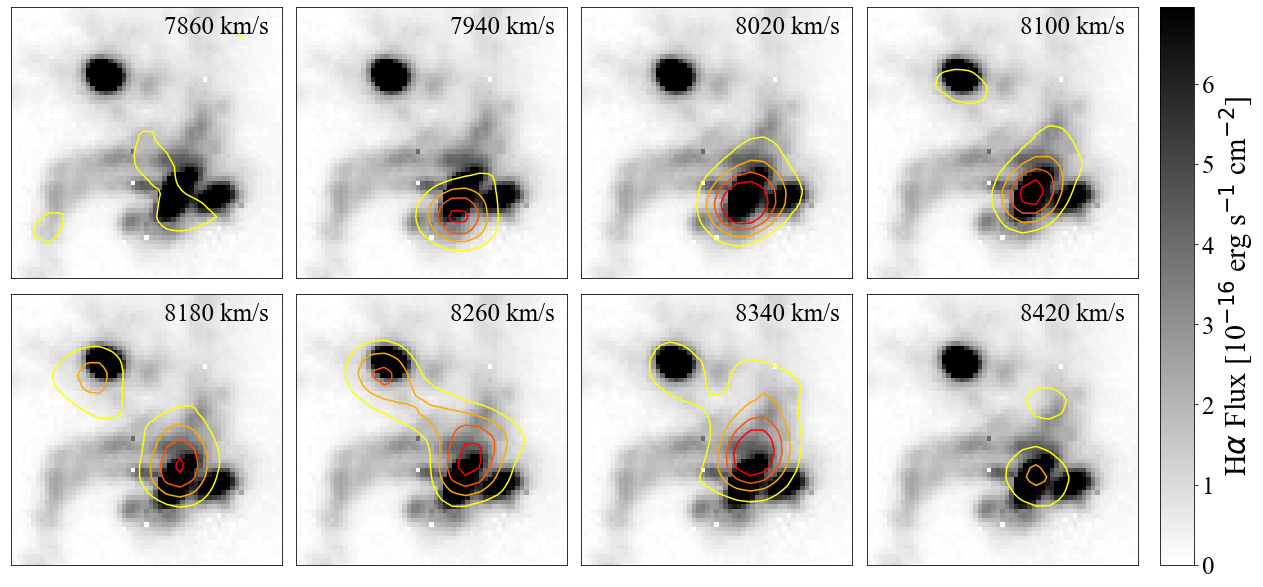}
    \caption{Distribution of the CO gas in the central region of Mrk~266 as a function of velocity from the CARMA datacube. Each panel is 19.2" by 19.2" and contains a grayscale image of the H$\alpha$ flux as observed by SITELLE and the CO emission at a fixed velocity, the contours indicate values of 0.02, 0.04, 0.06 and 0.08 Jy/beam as observed by CARMA.
    \label{figure:COcontours_on_Ha}}
\end{figure*}

The $^{12}$CO\,(1-0) emission at 115.3 GHz was observed for a total of 68 hours, but only four hours were used for the work presented here due to poor weather and technical issues. The data were acquired with the Combined Array for Research in Millimeter-wave Astronomy (CARMA) at the Owens Valley Radio Observatory. The data reduction and analysis for Mrk~266 (and 29 other nearby LIRGs) are detailed in Alatalo et al. (in prep) and identical to the reduction described in \citep{Alatalo2013}.  The CARMA correlator setup allowed us to detect gas in the 7916–8762 km/sec velocity range. The $^{12}$CO\,(1-0) flux map is the sum of the 7860~km/s to 8420~km/s channels. The spatial distribution of the CO emission as a function of its velocity is shown on Figure~\ref{figure:COcontours_on_Ha} and at velocities between 8100 km/s and 8340 km/s, CO is detected both in the nuclei as well as in between the two nuclei. Flux is recovered over a field of view of 1'x1.5' but the region where detections are above 3$\sigma$ is  $\sim$15"x20". The final synthesized beam major and minor axis are respectively 3.9$^{\prime \prime}$ and 3.3$^{\prime \prime}$, with a position angle (PA) of 77.11$^\circ$ and a detection limit of 1.3~Jy/beam~km/sec. At a distance of 129~Mpc, an angular size of 1$^{\prime \prime}$ corresponds to $\sim$0.6~kpc, so the CO data corresponds to a physical scale resolution of $\sim$2~kpc. The total CO flux was $\sim$230~Jy~km/sec.\\

We use the Fourier transform imaging spectrometer  \cite[SITELLE, ][]{Drissen2019} on the Canada-France-Hawaii Telescope, on May 11 and 12, 2019, to map the ionized gas emission H$\alpha$($\lambda6563$\AA), [NII]$\lambda \lambda 6549,6583$\AA, H$\beta$($\lambda4861$\AA), and [OIII]$\lambda\lambda 4959,5007$\AA. 
The pixel size is 0.32" but the average seeing is 0.8". SITELLE’s impressive field of view ($11^\prime \times 11^\prime$) allowed us to map for the first time in a spectral line the large scales, extended emission previously only observed with wide or at best narrowband filters. The data cubes were reduced, and emission line fits were performed using dedicated Python modules tailored for SITELLE \citep[ORBS and ORCS,][]{Martin2017, Martin2021}. Spectra in several regions throughout the galaxy have been checked by hand and did not show evidence of significant stellar absorption, thus no correction for absorption is made on the basis that the observed wavelength domain do not cover many stellar features that can be used to study and subtract the absorption component underneath H$\alpha$. This paper only considers the H$\alpha$ emission extracted from the SN3 filter (647-685~nm with R~=~2000) and while two velocity components along different sightlines are visible, we present the total emission here. An extended study of the SITELLE dataset will be presented in Petric~et~al.(in~prep).\\

Figure~\ref{figure:COcontours_on_Ha}, shows that the CO flux centered near the two nuclei, with most of the CO surrounding the SW nucleus. A bridge of CO emission connects the two nuclei. The H$\alpha$ emission extends over the whole system, displaying high dynamic range filaments and tidal tails, with high-intensity peaks at the two nuclear regions and in the Northern loop.


\section{Results} \label{results}

\subsection{Regions of interest} \label{section:regions}
\begin{figure}
    \centering
    \includegraphics[width = \columnwidth]{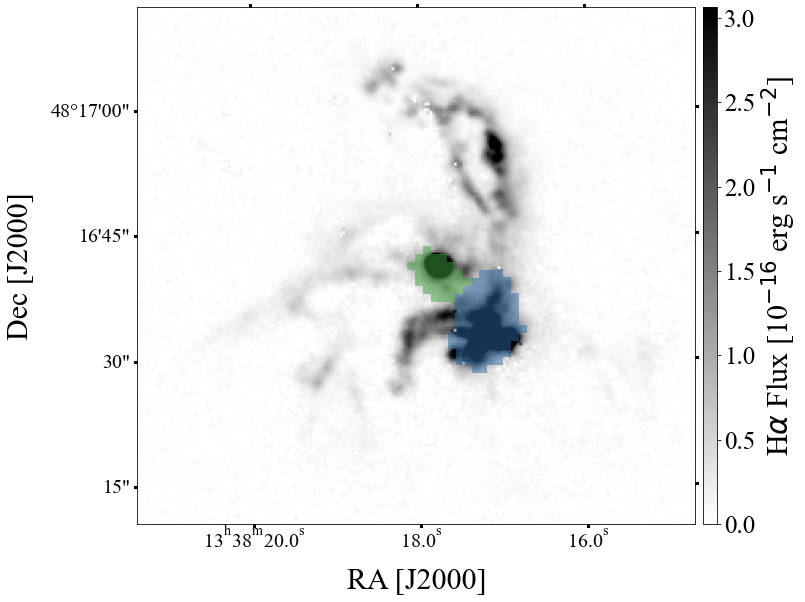}
    \caption{Defined regions of interested in Mrk~266 superposed on a grayscale image of the H$\alpha$ emission. The region in blue indicates the SW region, the green region refers to the NE region and the combination of these two regions forms the central region of Mrk~266.
    \label{figure:COregions_on_Ha}}
\end{figure}

Due to the heterogeneous dataset, the resolution of the ionized gas (SITELLE, 0.8") is more precise than the resolution of the molecular gas (CARMA, 3.9"~x~3.3"); only total measurements of particular regions are compared and no pixel-by-pixel comparison is made. We define the central region of Mrk~266 as all pixels within a 3$\sigma$ cut on the CO flux map. This region is also divided by a diagonal line to distinguish the two ellipses that defines the SW and NE regions (colored blue and green on fig.~\ref{figure:COregions_on_Ha}).\\

BPT diagrams indicate that most of the SITELLE pixels over the whole systems have some shock signature (i.e they fall in the composite region of the [OIII]/H$\beta$ vs [NII]/H$\alpha$ BPT diagram; Petric~et~al.~in~prep.). \cite{Wang1997} computed electron temperatures from the [OIII]$\lambda4363$/[OIII]$\lambda\lambda4959,5007$ and [NII]$\lambda5755$/[NII]$6548,6584$ emission lines ratios in the NE and SW regions and concluded that photoionization is the dominant heating mechanism. \cite{Wang1997} also states that a starburst can not be the entire ionization source and a hard radiation field of an AGN is required.

\subsection{CO Velocity} \label{COtovelocity}
\begin{figure}
    \centering
    \includegraphics[width = \columnwidth]{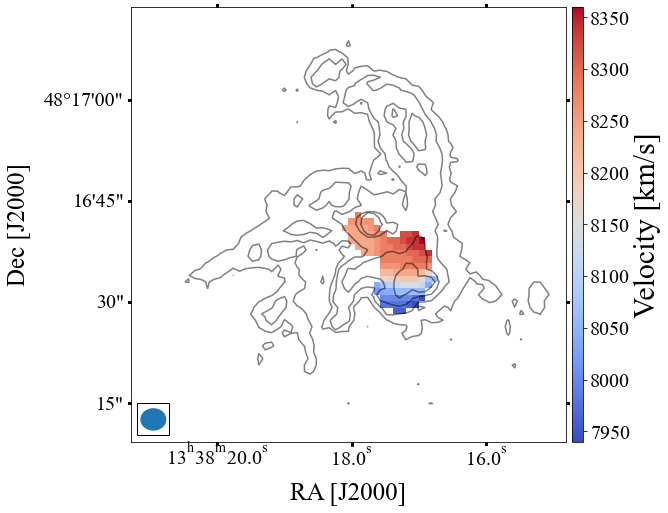}
    \caption{CO velocity map of Mrk~266. The measured CO systemic velocity is 8460~km/s with a velocity resolution of 21 km/sec. Only pixels above the detection threshold of 3$\sigma$ are shown. H$\alpha$ flux contours are superposed to the CO velocity map (contours are for fluxes of 1.2, 7, 30 and 90 $\times10^{17}$ erg~cm$^{-2}$~s$^{-1}$). The CO beam (3.9$^{\prime \prime} ~\times ~3.3^{\prime \prime}$) is shown in the lower-left corner.
    \label{figvelCO}
    }
\end{figure}

To determine the velocity centroid of each CO emission line we fit a single component Gaussian for each pixel in the CO cube. Figure~\ref{figvelCO} displays H$\alpha$ emission line flux contours atop the cold gas velocity. The radial velocity appears uniform across the NE nucleus while the velocities across the SW system suggest rotation, and a possible stable cold gas disk.\\

We measure the velocity dispersion of the two nucleus regions separately \text{following the definition detailed in Section~\ref{section:regions}}. The CO line widths and, therefore, the CO velocity dispersion of these regions are estimated using the FWHMs of the Gaussian fits. We find a velocity dispersion of 393$\pm167$~km/s for the SW nucleus and 205$\pm87$~km/s for the NE nucleus.

\subsection{Dynamical Mass} \label{COtodyn}
We estimate the dynamical mass of the two nuclei by following the methods employed by \cite{Evans2005} and \cite{Boselli2011}, using
\begin{equation}
    M_{\rm dyn} = \frac{3}{2}\frac{\Delta v^2_\text{FWHM}R_\text{CO}}{G},\label{Mdyn_eq}
\end{equation}
where $\Delta v_\text{FWHM}$ is the CO velocity dispersion calculated in Section~\ref{COtovelocity}, G is the gravitational constant and $R_\text{CO}$, the radius of the CO distribution. This last parameter is given by:
\begin{equation}
    R_\text{CO} = \sqrt{\frac{L'_\text{CO}}{\pi \,T_\text{bb} \, \Delta v_\text{FWHM}}}, \label{Rco}
\end{equation}
where $L'_\text{CO}$ is the CO luminosity in K~km~s$^{-1}$~pc$^2$, and $T_\text{bb}$ is the black body temperature of the dust, set to 50~K as done by \cite{Mazzarella2012}.\\

Equation~\ref{Rco} assumes an optically thick and thermalized gas with a filling factor of 1. The CO velocity distribution of the NE nucleus is single-peaked, and that of the SW nucleus is double-peaked, suggesting different galaxy geometries. For simplicity and in the absence of a reliable geometry, we use the same constant of 3/2 in Equation~\ref{Mdyn_eq} for both nuclei. Ideally, Equations~\ref{Mdyn_eq} and \ref{Rco} require a correction to account for a possible influence of a nucleus on the kinematics but the data presented here do not suggest outflow and near-equilibrium is assumed. However, Mrk~266 is a complex system and outflows are observed in other lookalike system (i.e. intermediate mergers with dual AGN) like VV~114 \cite[]{Yun1994} and NGC~6240 \cite[]{Treister2020}. Higher resolution CO observation should constrain this concern. The dynamical masses calculated are 1.3$\pm0.8\times10^{10}$~$M_\odot$ and 2.1$\pm1.4\times10^9$~$M_\odot$ for the SW and NE nucleus regions, respectively. These masses are compiled in Table~\ref{table_mass&sfr}.\\

\begin{table*}
\caption{Masses and Star Formation Rates}
\label{table_mass&sfr}
\begin{tabular}{lccc}
\hline
& SW & NE & Total \\
\hline
$M_{\text{dyn}}$ ($M_{\sun}$) & $1.3\pm  0.8\times10^{10}$ & $2.1 \pm 1.4\times10^{9}$ & $1.5 \pm 1.4\times10^{10}$ \\
$M_{\text{H}_2}$ ($M_{\sun}$)& $9.9~\pm 2.3~\times10^{9}$ & $1.7~\pm 0.4~\times10^{9}$ & $1.1~\pm 0.3~\times10^{10}$ \\
$\Sigma_{\text{H}_2}$ ($M_{\sun}$ pc$^{-2}$) & $332 \pm 83$ & $168 \pm 42$ & $288 \pm72$ \\
SFR$_\text{KDM12}$ ($M_\odot$~yr$^{-1}$) & $12 \pm 4$ & $0.8 \pm 0.3$ & $13 \pm 3$\\
SFR$_\text{H$_2$,spirals}$ ($M_\odot$~yr$^{-1}$) & $11 \pm 6$ & $2 \pm 1$ & $13 \pm 7$\\
SFR$_\text{H$_2$,starbursts}$ ($M_\odot$~yr$^{-1}$) & $43 \pm 29$ & $8 \pm 5$ & $51 \pm 35$\\
SFR$_\text{H$_2$,general}$ ($M_\odot$~yr$^{-1}$) & $22 \pm 9$ & $3 \pm 1$ & $25 \pm 10$\\
SFR$_\text{H$\alpha$,M1}$ ($M_\odot$~yr$^{-1}$) & $ 10.0\pm1.0 $ & $4.8\pm0.5$ & $14.8\pm2.0$\\ 
SFR$_\text{H$\alpha$,M2}$ ($M_\odot$~yr$^{-1}$) & $7.0\pm0.7 $ & $3.4\pm0.3$ & $10\pm1$\\ 
SFR$_\text{H$\alpha$+IR,M3}$ ($M_\odot$~yr$^{-1}$) & $14.5 \pm1.5$ & $4.0\pm0.4$ & $18.5\pm1.9$\\ 
SFR$_\text{IR,Mur11}$ ($M_\odot$~yr$^{-1}$) & $56$ & $9$ & $65$\\ 
SFR$_\text{IR,C15}$ ($M_\odot$~yr$^{-1}$) & $33$ & $5$ & $38$\\ 
\hline
\end{tabular}
\end{table*}

\subsection{Molecular Hydrogen Mass from CO} \label{COtomass}
The following equation, from \cite{Evans2005}, links the CO flux density, $S_\text{CO} \, \Delta\nu$ (in Jy~km~s$^{-1}$), to the corresponding H$_2$ mass:
\begin{equation}
  M_{\text{H}_2} = \: \alpha_{\text{CO}}\left[2.4\times 10^3  \, S_\text{CO} \, \Delta\nu \, D_L^2 \, (1+z)^{-1} \right]\label{COtoH2} .
\end{equation}

We adopt a luminosity distance and redshift of $D_L$~=~129~Mpc and $z$~=~0.0279 \citep{Mazzarella2012}. We assume a CO-to-H$_2$ conversion factor of $\alpha_\text{CO}$\,=\,2.0\,$\pm0.5$\,$M_\odot$\,pc$^{-2}$\,(K\,km\,s$^{-1}$)$^{-1}$. This value is higher than the typical conversion factor of 0.6\,$\pm0.2$\,$M_\odot$\,pc$^{-2}$\,(K\,km\,s$^{-1}$)$^{-1}$ for LIRGs \cite[]{Bolatto2013}, but recent studies \citep{Narayanan2011, Narayanan2012, Cicone2018, Treister2020} suggest that the CO-to-H$_2$ conversion factor for mergers is a few times lower than the galactic average but higher than the typical LIRGs value. Simulations have shown that the conversion factor in mergers is highly dependent on the merging phase. However, a conversion factor of $\alpha_\text{CO} = 2$~$M_\odot$~pc$^{-2}$(K~km~s$^{-1}$)$^{-1}$ may be suitable for merger systems as long as they are not in the final coalescence phase \cite[]{Narayanan2011}. Our CO-to-H$_2$ conversion factor is also higher than the conversion factor of 1\,$M_\odot$\,pc$^{-2}$\,(K\,km\,s$^{-1}$)$^{-1}$  used by \citet{Bolatto2013} for starbursts, but lower than 3.2\,$M_\odot$\,pc$^{-2}$\,(K\,km\,s$^{-1}$)$^{-1}$ as used by \citet{Treister2020} for the dual AGN host and advanced merger in NGC~6240.\\

We use Equation~\ref{COtoH2} to derive an H$_2$ mass map ($M_\odot$ per pixel). We sum all the pixels above the 3$\sigma$ threshold to estimate a molecular gas mass of $M_\text{H$_2$}=1.1\pm0.3~\times10^{10}$~$M_\odot$ for the central region of Mrk~266 (identified as the Total region in Table~\ref{table_mass&sfr}). We include the emission in the nuclei and molecular gas bridge between them. To calculate the mass of each nucleus separately, we use the delineation described in Section~\ref{COtovelocity}. We compute gas surface density in physical units by using a scaling relation of 590~pc~per~arcsec to compute the physical area represented by each pixel, employing the distance provided to the system given in  \citet{Mazzarella2012}. We summarize those masses and surface density in Table~\ref{table_mass&sfr}.

\subsection{Star Formation Rates from H$_2$} \label{surftoSFR}
\subsubsection{Volumetric Star Formation Law} \label{sfr_kdm12method}

The amount of cold gas present can predict a future star formation rate limit. We use the method of \citet[][KDM12 hereafter]{Krumholz2012}, a technique that appears to be viable for many types of galaxies, from Milky Way-like to starburst galaxies. It considers the gas volume density $\rho$ and links the star formation rate to the molecular gas according to the local free-fall time ($t_{\rm ff}$). If we assume that the central region of Mrk~266 contains more hydrogen in the molecular phase than in the atomic phase \citep{Bigiel2008, Leroy2008, Kennicutt2021}, the SFR surface density, according to \citetalias{Krumholz2012}, is given by:
\begin{equation}
    \Sigma_{\text{SFR}} = f_{\text{H}_2} \epsilon_\text{ff}\frac{\Sigma_{\text{gas}}}{t_\text{ff}}. \label{KDM12}
\end{equation}

Here, $f_{\text{H}_2}$ is the mass fraction of molecular hydrogen given by $M_{\text{H}_2} / (M_{\text{H}_2} + M_*)$ where $M_{\text{H}_2}$ is the total molecular gas mass within a region as calculated in Section~\ref{COtomass}. $M_*$ is the total stellar mass within the nuclei, which we set to $6.1\times10^{10}~M_\odot$ and $4.4\times10^{10}~M_\odot$ for the SW and the NE regions respectively based on the work of \cite{Mazzarella2012}. The parameter $\epsilon_\text{ff}$ is dimensionless and relates the star formation per volume to the fraction of mass in molecular form, and the free-fall time. This factor is fairly constant though its been shown to have a slight functional dependence on the virial ratio and Mach numbers, in star-forming region its value is close to 0.015 according to \cite{Krum2005,Krumholz2012,Krumholz2013}. Equation \ref{KDM12} assumes that star-formation takes place in filamentary giant molecular clouds, observed at random orientation, and thus effectively at the resolution of most extragalactic observation are well described as spherical structures, with density and filling factors believed typical for starburst galaxies in the local universe \citep{Kennicutt1998}.\\

The ratio ${\Sigma_{\text{gas}} / t_\text{ff}}$ (often written as $[\Sigma_{\text{gas}}/t]_\text{single-ff}$) describes the relation between the gas surface density and the averaged single free-fall time. The gas surface density was calculated in Section~\ref{COtomass}. The free-fall time is given by the equation: $t_\text{ff}=(3\pi/32G\rho)^{1/2}$, where $G$ is the gravitational constant and $\rho$, the average gas density. \cite{Federrath2017} defines the density as  $\rho=\overline{\Sigma_{\text{gas}}}/L$ with $\overline{\Sigma_{\text{gas}}}$, the average gas surface density and $L=\text{H}/\cos\theta'$, H is the scale height which is believed to be 10~pc in ULIRGs \cite[]{Krumholz2012} and $\theta'=90^\circ-\theta$, where $\theta$ is the galaxy inclination. \cite{Iwasawa2020} measured an inclination of $72^\circ$ for the SW nucleus. We do not have an inclination for the NE galaxy and adopt $\theta$ = 90$^\circ$ for this region. The SFR estimates, for the individual nucleus regions (as defined in Section~\ref{COtovelocity}) and for the total central region of Mrk~266 are compiled in Table~\ref{table_mass&sfr}. We find that the central SW has the highest rates on the order of $12 \pm 4 ~\rm{M}_\odot$~yr$^{-1}$ while the NE region shows significantly lover values at $0.8 \pm 0.3 ~\rm{M}_\odot$~yr$^{-1}$.\\

It is important to note that this method assumes a scale height of 10~pc and that the gas within the selected region is a self-gravitating system. These caveats lessen the reliability of this method when applied to a complex system like Mrk~266. However, these results are still relevant when compared to the other methods used to estimate the probable future star formation.

\subsubsection{Schmidt-Kennicutt Law} \label{sfr_k21method}
\cite{Schmidt1959} and \cite{Kennicutt1998} have established a linear relation between $\Sigma_\text{H$_2$}$ and $\Sigma_\text{SFR}$, the surface density of molecular gas and the present-day SFR based on IR, respectively. \cite{Kennicutt2021} presents an updated version of the classic Schmidt-Kennicutt power law using galaxies over a broad range of surface density. The power laws fitted by \cite{Kennicutt2021} for normal spirals, starburst galaxies, and their whole sample are:
\begin{equation}
    \Sigma_\text{SFR, spiral} = 10^{(-3.76 \pm 0.08)}\:\Sigma_\text{H$_2$}^{(1.34 \pm 0.07)}, \label{k21_spirals}
\end{equation}
\begin{equation}
    \Sigma_\text{SFR, starburst} = 10^{(-2.27 \pm 0.21)}\:\Sigma_\text{H$_2$}^{(0.98 \pm 0.07)} .\label{k21_starbursts}
\end{equation}
\begin{equation}
    \Sigma_\text{SFR, general} = 10^{(-3.87 \pm 0.04)}\:\Sigma_\text{H$_2$}^{(1.54 \pm 0.02)}, \label{k21_general}
\end{equation}
where $\Sigma_\text{H$_2$}$ is in units of
$M_\odot$~pc$^{-2}$ and $\Sigma_\text{SFR}$, in units of 
$M_\odot$~yr$^{-1}$~kpc$^{-2}$. The SFRs calculated using these power laws are presented in Table~\ref{table_mass&sfr}.
The uncertainties are determined by propagating the uncertainty on $\Sigma_\text{H$_2$}$ and the uncertainties within the power laws.

\subsection{Star Formation Rate from H$\alpha$}\label{SFRfromHa}
Estimating star formation rates in sources with multiple, interacting sources of photoionization and shocked gas is challenging. Considerable observational and theoretical efforts allow us to estimate the star-formation rates using H$\alpha$ and hybrid methods using H$\alpha$ and infrared emission \cite{cat2015, Calzetti2000,als2006, ken2009, ds2010}. We estimate the star-formation rates using three comparable methods: (1) uses \cite{Kennicutt1998}, (2) use a method developed to estimate the SFR rates of type 2 AGN from the extinction corrected H$\alpha$ emission \cite[]{cat2015}, and (3) use a hybrid IR and H$\alpha$ emission calibrated for type 2 AGN \cite[]{cat2015}. We select only pixels that have a CO counterpart and reject the pixels at the positions of the two X-ray emission peaks associated with the AGN \citep[as given by][]{Mazzarella2012} to avoid AGN contamination.

\subsubsection{Method 1} \label{sec:method1}
Considering a continuous mode of star formation with a standard initial mass function, \cite{Kennicutt1998} relates the luminosity of the integrated H$\alpha$ emission over a galaxy to its present-day SFR: 
\begin{equation}
    \text{SFR}\:[M_\odot \text{yr}^{-1}] = 7.9 \times 10^{-42} L(\text{H$\alpha$})\:[\text{erg}\:\text{s}^{-1}]. \label{HaK98}
\end{equation}
The H$\alpha$ emission of Mrk~266 was corrected for internal extinction assuming using the Balmer decrement,  
The extinction due to the Milky Way is assumed negligible based on the data of \citet{Schlafly2011}.
\newline

We estimate a total SFR for the central region of Mrk~266 of $14.8\pm1.5~M_\odot$~yr$^{-1}$. The uncertainty given here includes a 10\% uncertainty on the H$\alpha$ flux. Using the masks described in Section~\ref{COtovelocity} to isolate the nucleus regions, we calculated the SW and NE SFR$_{{\rm H}\alpha}$ separately. Values are given in Table~\ref{table_mass&sfr} as SFR$_\text{H$\alpha$,M1}$. Due to the difference in size between the SW and the NE regions (as defined in Section~\ref{section:regions}), the total SFR of the SW region is higher than the NE (Table~\ref{table_mass&sfr}) but its SFR surface density reveals that it is less dense than the NE (Figure~\ref{K21comp}).

\subsubsection{ Method 2}
Spatially resolved, spectroscopic studies of nearby galaxies, including AGN hosts, combined with multi-wavelength IR-through radio were used to develop SFR diagnostics based on optical and IR emission for normal galaxies and also galaxies hosting type 2 AGN \cite[]{cat2015}. 

To compute the attenuation correction using the Balmer decrement, we assume a foreground screen dust geometry. Such simple dust geometry is unlikely for this complex system but without detailed knowledge of the spatially resolved MIR spectral properties it is not feasible to develop a more realistic dust model. To correct the H$\alpha$ emission for dust extinction, we assume that the intrinsic Balmer decrement is 2.86 for case B recombination, at electron temperature T$_{e}~ =~ 10,000~K$ and use the H${\alpha}$ and H${\beta}$ extinction coefficients from \citet{card1989}: 2.53 and 3.61 respectively. This approach is similar to that taken by \cite{cat2015}. Using their conversion from extinction corrected H$\alpha$ luminosity, L$(\rm{H\alpha}_{\rm{cor}})$, to SFR (equation~\ref{HaC15}) we find a SFR of $\sim 10\:M_\odot \text{yr}^{-1}$ for the SW nucleus and $\sim 5\:M_\odot \text{yr}^{-1}$ for the NE nucleus. 

\begin{equation}
    \text{SFR}\:[M_\odot \text{yr}^{-1}] = 5.5 \times 10^{-42} L(\text{H$\alpha$}_{\rm{cor}}). \label{HaC15}
\end{equation}

 
\subsubsection{ Method 3 }
\citet{cat2015} find that a combination of H$\alpha$ and IR give consistent SFR rates for type 2 AGN hosts, even when Balmer decrement measurements are not available. We use their formalism (equation \ref{HaC15t2}), the observed H$\alpha$ luminosity measured from the SITELLE data (L$(\text{H$\alpha$}_\text{obs})$), and the Herschel photometry and 70$\mu$m maps of \citet{chu2017} to estimate an SFR of $\sim$14.5~$M_\odot~\text{yr}^{-1}$ for the SW nucleus and $\sim$4~$M_\odot~\text{yr}^{-1}$ for the NE nucleus (labeled as SFR$_\text{H$\alpha$+IR,M3}$ in Table~\ref{table_mass&sfr}). 

\begin{equation}
    \text{SFR}\:[M_\odot \text{yr}^{-1}] = 5.5 \times 10^{-42} [L(\text{H$\alpha$}_\text{obs})~+~ a_{IR} \times L_{IR}]. \label{HaC15t2}
\end{equation}

The IR luminosity $L_{IR}$ is estimated from the 70$\mu$m photometry using the formalism of \cite{zak2016,sym2008,petric2015}. The $a_{IR}$ coefficient for the calibration of hybrid (i.e. using both $L\text{H$\alpha$}$ and $L_{IR}$) SFR tracers is $\sim$0.0015 for galaxies hosting a type 2 AGN \cite{cat2015}.


\subsubsection{Comparison methods based solely on IR observations}
The MIR and IR spectral and photometric properties of Mrk~266 have been detailed in \citep{chu2017, Mazzarella2012,Petric2011,stierwalt2014, inami2013, Petric2018, Howell2010}. Here, we use the Herschel PACS images (pixel size of 3.2") 
and photometry from \citet{chu2017} to estimate the relative IR luminosity of the two nuclei and star-formation estimates.  Again, because of the heterogeneous resolutions of the dataset, no attempt is made to compare SFR from different indicators pixel-by-pixel. 

Using the conversion from IR to SFR in \citep{Murp2011} for normal galaxies gives us a higher values ($\sim$56~$M_\odot~\text{yr}^{-1}$) than the value ($\sim$33~$M_\odot~\text{yr}^{-1}$ ) we obtain when using the formalism of \cite{cat2015} for type 2 AGN, for both nuclei. MIR continuum emission in galaxies arises from a combination of ionized interstellar gas, evolved stellar population, non-thermal emission from radio sources, very small grains, and PAHs. We can estimate the AGN contribution to the MIR emission in individual galaxies from the ratios of high to low ionization fine-structure emission lines. Dust continuum and dust features provide additional diagnostics of the relative contribution of starburst and AGN to the MIR luminosity. Observations and theoretical models show that the AGN radiation field can heat dust grains to such temperature that the dust continuum emission becomes prominent between 3 and 6$\mu$m. Galaxies with an AGN tend to have low 6.2$\mu$m PAH equivalent width (EQW) due to the presence of a significant hot dust continuum and because the hard AGN photons may destroy the PAH molecules \citep[]{armus2009, Petric2011, stierwalt2014}. Spitzer MIR spectra of the NE and SW component reveal that the PAH EQW of the NE component is 0.4$\pm$0.1 $\mu$m, smaller than that of the SW component 0.67$\pm$0.03 $\mu$m \cite{Mazzarella2012}. Both values are close to the typical value of 0.54 $\mu$m for star-forming galaxies \citep{Smith2007}. The relatively lower 6.2$\mu$m PAH EQW may suggest that the PAHs around the NE component are more impacted by the AGN than the PAHs around the SW component. This is consistent with our finding that the SW component has more CO and may have its molecular gas shielded from the AGN radiation by a high-obscuring column density of dust. This is also consistent with the \cite{Iwasawa2020} discovery that the SW component is a Compton thick AGN.

\section{Discussion}
\label{discussion}
We estimate that the total molecular gas reservoir in the central region of Mrk~266 is $1.1\pm0.3\times10^{10}$~$M_\odot$. Using the regions defined in Section~\ref{section:regions}, the SW and NE regions are expected to harbor $9.9\pm2.3\times10^{9}$~$M_\odot$ and $1.7\pm0.4\times10^{9}$~$M_\odot$ respectively. These estimations are based on  observation of the $^{12}$CO (1-0) emission by the CARMA observatory with a beam size of 3.859" by 3.335". Single dish observations of Mrk~266 by \cite{Sanders1986} estimate a total H$_2$ mass of $1.4\times10^{10}$~$M_\odot$ within a 45" beam. \cite{Imanishi2009} carried out CO ($\nu_\text{rest} = 115.271$ GHz), HCN(1–0) and HCO$^+$(1–0) observations using the Nobeyama Millimeter Array (NMA) with a beam size of 4.3" by 3.4". The beam size of the \cite{Sanders1986} observations did not allow to distinguish individual emission of the nuclei but both the NMA observations presentented in \cite{Imanishi2009} and the CARMA observations presented in this work allows to do so. \cite{Mazzarella2012} estimated the H$_2$ mass of the nuclei as $3.4\times10^{9}$~$M_\odot$ for the SW, $7.0\times10^{8}$~$M_\odot$ for the NE and a total of $7.0\times10^{9}$~$M_\odot$ in Mrk~266 from the \cite{Imanishi2009} CO observations. These results assumed a CO-to-H$_2$ conversion factor of $1$~$M_\odot$~(K~km~s$^{-1}$~pc$^2)^{-1}$ while the results of this work assume $2$~$M_\odot$~(K~km~s$^{-1}$~pc$^2)^{-1}$ as justified in Section~\ref{COtomass}. These high resolution interferometric observations also indicate that there is an offset between the CO emission and the H$\alpha$ emission peaks (see Figure~\ref{figure:COcontours_on_Ha} and \ref{figvelCO}). \\ 

\begin{figure}
    \centering
    \includegraphics[width = 0.9\columnwidth]{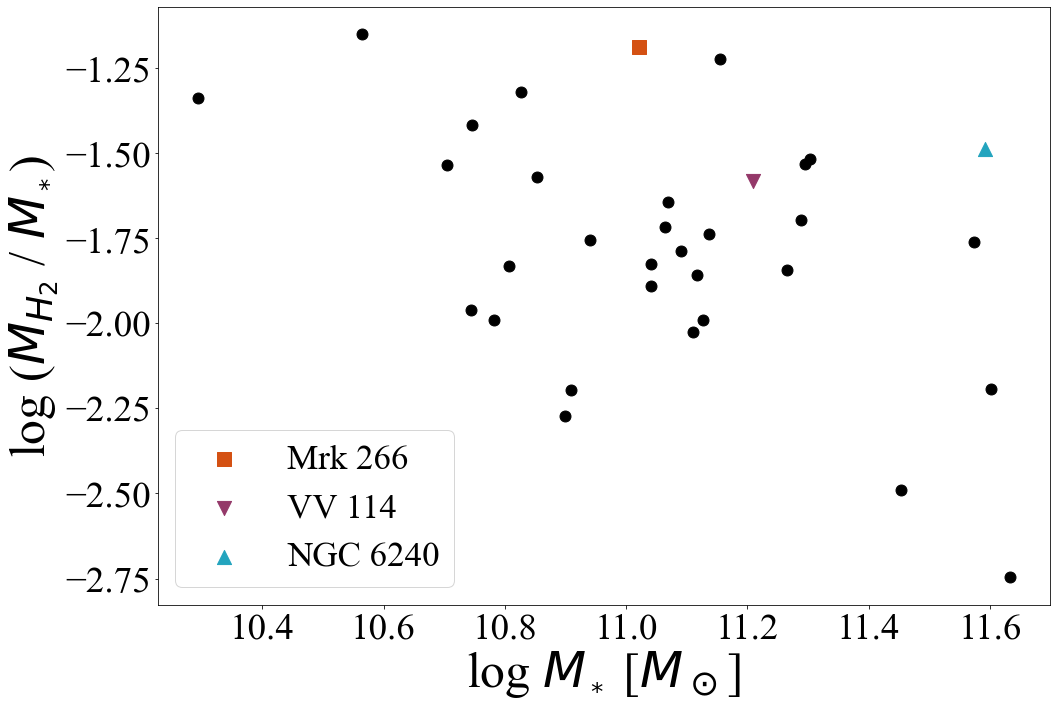}
    \caption{Comparison of molecular masses and stellar masses of nearby luminous IR galaxies. Black points are galaxies that are both present (based on a name correspondence) in  \protect\cite{Yamashita2017} for  $M_{H_2}$,  and \protect\cite{Howell2010} for $M_*$. Mrk~266 is represented by an orange square, using our result for $M_{{\rm H}_2}$ while its $M_{*}$ is taken from \protect\cite{Mazzarella2012}. VV~114, considered in a similar evolutionary stage as Mrk~266, is represented by a downward violet triangle. NGC~6240, thought to be in a slightly more advanced evolutionary stage is represented using a blue upward triangle from \protect\cite{Treister2020} results.}
    \label{MH2vsM}
\end{figure}

Figure~\ref{MH2vsM} shows how the gas and stellar mass of Mrk~266 compares to those of other nearby LIRGs using the stellar masses of \citet{Howell2010} and single-dish estimates of molecular gas masses from \citet{Yamashita2017}. Mrk~266  follows the trend indicated by these galaxies. NGC~6240, another LIRG and dual AGN host, has higher SFR to H$_2$ ratios likely because it is a more advanced merger \cite[]{Treister2020}, with more compact dust ISM which can shield the molecular gas from the AGN radiation. The LIRG VV~114 (violet triangle) is an intermediate evolutionary merger stage LIRG,  similar to Mrk~266 \cite[]{Mazzarella2012} and its position in the diagram is also near the location of Mrk~266. Therefore, Mrk~266 displays typical fluxes and activity for its evolutionary phase relative to other LIRGs.\\

We measure dynamical masses from the CO velocity dispersion. The value obtained (see Table~\ref{table_mass&sfr}), are consistent with \cite{Mazzarella2012} results. As expected, dynamical masses are higher than these estimated from the CO luminosities. Furthermore, theory \cite[]{li2018}, and observations \cite[e.g., NGC\,7479][]{Fadda2021} find that as much as 50\% of the H$_2$ mass may lie in CO-dark regions because of low metallicity gas. The presence of CO-dark gas means that masses and star formation rates estimated from CO luminosities may be lower limits to the actual star-formation rates. The presence of low-metallicity gas in the center of a galaxy merger may be surprising but observations \citep{Kewley2006a,kew2010,Rupke2008} and simulations \citep{Montuori2010, Rupke2010a} have confirmed that galaxy mergers exhibit flatter metallicity gradient than isolated spiral galaxies. \citet{Kewley2006a,kew2010} proposed a scenario where merger dynamics dilute the central metallicities through large gas flows towards the central region. The SITELLE data show complex velocity structures, which may be associated with such gas inflows and \citet{hut1988} find 21~cm absorption against the NE nucleus possibly due to infalling gas. While the CO data presented here cannot accurately estimate the amount of dark CO, we can use the dynamical CO mass estimates to compute an absolute upper limit for the star formation rates. We use the starburst power law presented in Section~\ref{sfr_k21method} and the dynamical masses (Section~\ref{COtodyn}) to find SFRs of 61 $M_\odot$~yr$^{-1}$ and 10 $M_\odot$~yr$^{-1}$ for the SW and NE region respectively.\\

\begin{figure}
    \centering
    \includegraphics[width = \columnwidth]{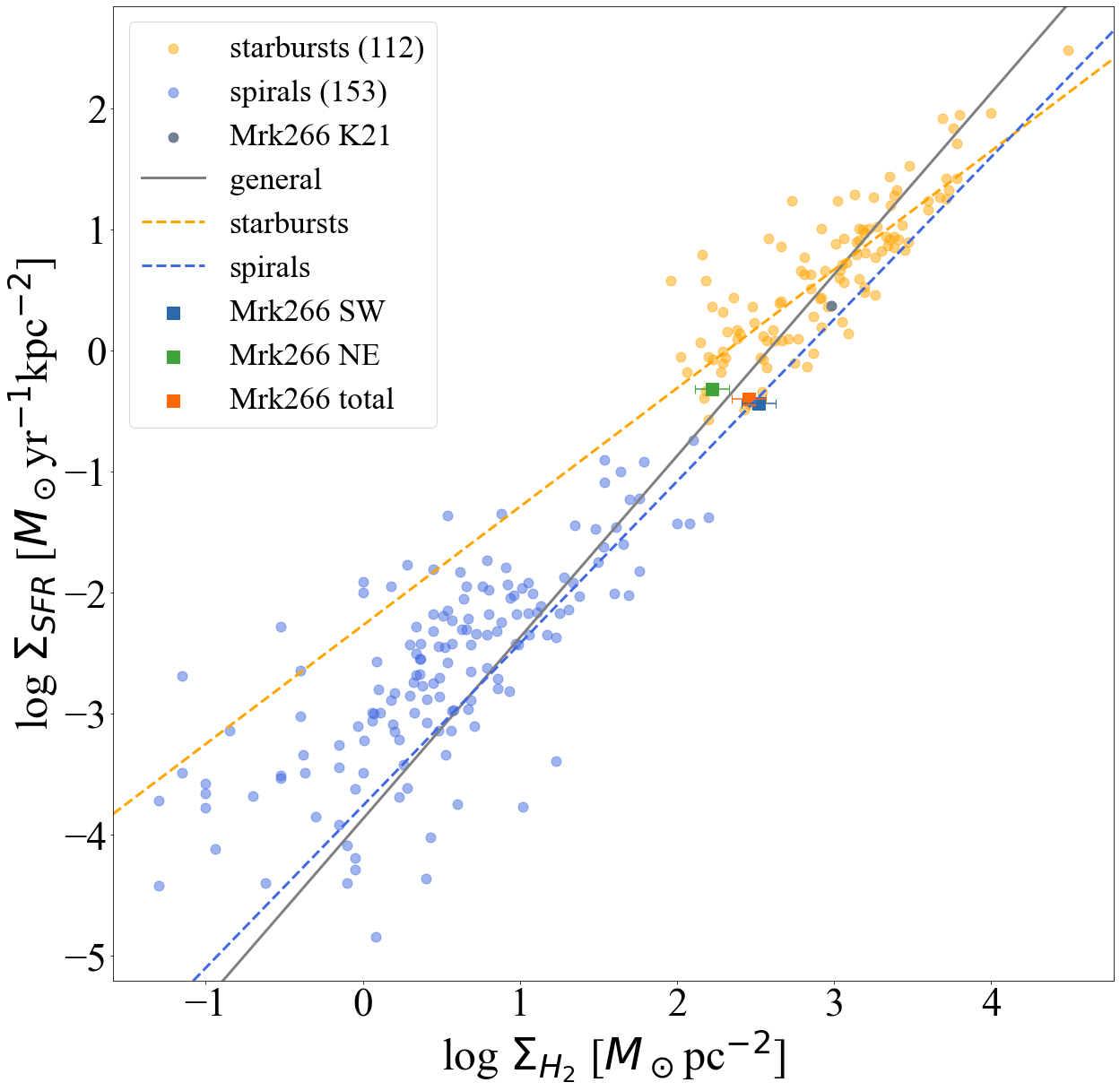}
    \caption{Comparison of the molecular and SFR$_{\rm IR}$ surface density using the data of \protect\cite{Kennicutt2021} (for starburst galaxies, in orange) and \protect\cite{delosReyes2019} (for spiral galaxies, in blue) with the addition of Mrk~266. Mrk~266 $\Sigma_{\text{H}_2}$ are determined from our CO data (Section~\ref{COtomass}) and $\Sigma_{\rm SFR}$ are from our H$\alpha$ data for the corresponding CO regions (Method~1, Section~\ref{sec:method1}). Different colors are used to identify Mrk~266 nucleus regions, as indicated on the plot. The solid gray dot is also Mrk~266 as positioned by the \protect\cite{Kennicutt2021} using the IR Luminosity from  \protect\cite{armus2009} and the CO measurement from \protect\cite{Sanders1991}.}
    \label{K21comp}
\end{figure}

Assuming that the relation between the H$\alpha$ and IR luminosity is close to 1 \cite[]{Kewley2002}, we plot our results over \cite{Kennicutt2021} data (Figure~\ref{K21comp}) and our measurements are within the scatter of their relation and therefore, our SFRs estimates from the uncorrected H$\alpha$ (Method~1, Section~\ref{sec:method1}) are not unrealistic. \cite{Hayward2014} find that SFR estimates based on IR luminosity may be overestimated even in the absence of an AGN because of the contribution from an old stellar population which could mean that Mrk~266 current starburst will be followed by quenching. This may also explain why our measurements for Mrk~266 are in the lower limit of \cite{Kennicutt2021} starburst sample. \\

Mrk~266 is a rare nearby LIRG hosting two observable AGN. As such it is interesting to see if its star-formation properties differ significantly from those of other local starbursts. Figure~\ref{K21comp}, adapted from \cite{Kennicutt2021}, compares IR and H$_2$ surface density SFR estimates for normal and starburst galaxies. Mrk~266 is part of this sample (gray~dot), its IR based SFR surface density is an order higher than the H$\alpha$ SFR surface density presented in this study. However, this measurement is for the entire system and not limited to the central region of Mrk~266. The CO measurement allows us to predict the future star formation activity in the system based on the method of \citetalias{Krumholz2012} and \cite{Kennicutt2021}. The molecular mass observed suggests that the system will be able to sustain the star formation activity in the near future, i.e. the star formation rate estimated using molecular masses (with \cite{Kennicutt2021} general power law, Section~\ref{sfr_k21method}), is 25$\pm$10~M$_\odot$~yr$^{-1}$. Those values are below the values associated with a major starburst phase where H$_2$ surface densities can reach 10$^4$~M$_\odot$~pc$^{-2}$ with SFRs\,$\simeq$\,10$^2$~M$_\odot$~yr$^{-1}$; such high values are typically associated with late stage mergers (i.e. right before the coalescence of the SMBHs) but not as common in intermediate phase mergers like Mrk~266 \citep[]{sanders1996,Sanders2003, Ellison2013b,Koss2019,Kennicutt2021}.\\

We find significant differences between the two nuclear regions. The SW nucleus appears to have five times more molecular gas than the NE nucleus. The SW has a more extended star-forming activity with a higher SFR, while the NE is more compact, with a smaller SFR and a higher SFR surface density. The SW nucleus hosts a Compton Thick AGN, is highly obscured, and the ISM surrounding it is likely clumpy. Comparatively, the NE nucleus with associated H{\small I} absorption which indicates inflowing gas \cite[]{hut1988} may contain CO dark regions \citep[i.e., areas where the CO emission does not trace the full H$_2$ reservoir;][]{Wolfire2010,Bigiel2020}. If the inflowing gas has relatively low metallicity, it can contain less CO than normal ISM and the presence of significant CO-dark gas would mean that analysis based only on CO emission line luminosities alone underestimates the amount of molecular gas available for star formation. \\

\begin{figure}
    \centering
    \includegraphics[width = 0.75\columnwidth]{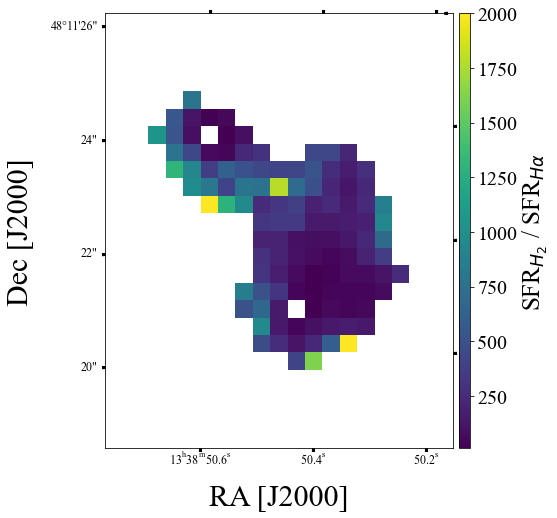}
    \caption{Ratios of the star formation rates from the cold gas and the ionized gas. Only pixels near the galaxy nuclei are shown, where the CO emission is over the 3$\sigma$ limit, but avoiding the AGN regions \citep[based on the X-ray peaks of][]{Mazzarella2012}.}
    \label{H2vsHa}
\end{figure}

We compare Mrk~266 present-day and future star formation rates and speculate that the star formation may be sustained and suggest that the system has recently experienced a massive inflow of gas. The gas inflow may have caused a starburst episode that may decrease due to a spike in the AGN activity or massive star activity (e.g., winds). Based on the total amount of molecular gas available to fuel future star formation and depending on the chosen H$_2$ based SFR (see Table~\ref{table_mass&sfr}), we speculate that the central region of Mrk~266 will sustain star formation for at least 215~Myr to 845~Myr at most. \\

The NE AGN (LINER) has a lower bolometric luminosity than the SW AGN \cite[Seyfert 2;][]{Iwasawa2020}. This difference is consistent with the idea that the NE AGN may quench the surrounding star-forming activity, but the high obscuration around the Compton thick AGN in the SW \cite[]{Iwasawa2020} prevents quenching. The CO emission in the SW nucleus appears to be in stable, smooth, velocity structures (Figure~\ref{figvelCO}). The H$_2$ column density we estimate assuming a spherical and homogenous geometry are significantly lower than the density of the gas and dust obscuring measured from the X-ray observations \citep{Iwasawa2020} suggesting highly clumpy or at least highly heterogeneous material. Such inhomogeneity shield the ISM (and star formation) from harsh AGN radiation and increase the momentum loading and drive potential outflows \citep{costa2014}.\\

Alternatively, an inflow of cold gas from the NE galaxy may enter the inner disk of the SW galaxy and sustain its star formation and the AGN activity. \citet{hut1988} find H{\small I} absorption in the Northern nucleus and suggest it may be associated with inflows that may fuel the AGN. The observed H{\small I} absorption in the SW nucleus is less significant. The observed molecular gas bridge between the nuclei and its velocity structure (Figure~\ref{figure:COcontours_on_Ha} and Figure~\ref{figvelCO}) suggests a possible smooth connection of the radial velocity between the NE component and the SW rotating disk. The idea that the SW galaxy accretes gas from the NE galaxy, fueling both the AGN and star formation is intriguing and needs to be confirmed with higher spatial and velocity data.\\

Theoretical and observational work indicate that the energy released by growing SMBHs is sufficient to affect the ISM of its host and even the diffuse gas in dark matter halos. AGN can quench and also enhance star formation by injecting energy into the gas, expelling it from the galaxy, or compressing dense clouds if the AGN-induced pressure is larger than the dynamic pressure controlling the ambient ISM \citep{Hopkins2006,Hopkins2008,silk2009,zinn2013,morganti2017}. The laboratory that is Mrk~266 provides for the study of how two AGN may shape the evolution of the galaxy and reveal a complex set of processes that allow both for vigorous star formation as discussed in this paper and expel materials at distances of more than 20~kpc \citep[e.g.][]{Mazzarella2012}. The molecular and ionized gas data we present here are consistent with an intermediate interaction phase for Mrk~266  \citep[]{Koss2012, Koss2019}. The two IR-luminous nuclei have different star-forming and AGN activities. Mrk~266 is a gas-rich merger system with plumes and arches of ionized gas as shown in Figure~\ref{image_Mrk266}, and it is not surprising that different AGN-connected processes affect the ISM on different scales. Higher resolution, sensitive molecular gas observations are needed to search for flowing gas as found in similar systems \citep{ala2019, aud2019,combes2019} and to test if the ISM is indeed clumpy around the nuclei of Mrk~266 as it is the case for other AGN host \cite[]{Treister2020}. Such observations could constrain the time scales of dual AGN and separate the impact of tidal interactions and AGN outflows for the enhanced obscured activity in the SW nucleus.

\section{Conclusion}
\label{conclusion}
We present estimates of the molecular gas mass and future star formation rates based on CARMA $^{12}$CO (1-0) observations in dual AGN host, and gas rich-merger Mrk~266. We also present estimates of the current star formation based on H$\alpha$ observations of Mrk~266 made with SITELLE at the CFHT. The data confirm that Mrk~266 is indeed an intermediate stage merger with an active and sustained star formation for the next few hundreds Myr. The two main galactic components appear to have vastly different activity levels; while the NE system passed its peak in star formation and SMBH growth, the SW is a highly obscured Compton thick AGN that seems to be surrounded by abundant and star-forming CO. We speculate that these differences may originate in contrasting boundary conditions that caused the SW interstellar medium to become clumpier and dustier or in merger geometry or even in a different gas to star ratio.\\
 
Higher spatial and spectral resolution observations of the CO line emission are needed to estimate the cold gas distribution and motion better. A study of the ionized gas emission lines in the visible with SITELLE is underway to estimate the recent star formation, the gas metallicity, dynamics, and the different ionizing sources. Ionized and molecular gas observations matched in resolution and sensitivity will provide additional clues needed to understand Mrk~266 and a unique opportunity to understand galaxy interactions at a merger stage where both AGN are present and detectable.


\section*{Acknowledgements}
The authors whish to thank the anonymous referee for the valuable comments which have significantly improved the paper. This work is based on observations obtained with SITELLE, a joint project of Université Laval, ABB, Université de Montréal, and the Canada-France-Hawaii Telescope which is operated by the National Research Council of Canada, the Institut National des Sciences de l’Univers of the Centre National de la Recherche Scientifique of France, and the University of Hawaii. The authors wish to recognize and acknowledge the very significant cultural role that the summit of Mauna Kea has always had within the indigenous Hawaiian community. We are most grateful to have the opportunity to conduct observations from this mountain. This research made use of Astropy, a community-developed core Python package for Astronomy \citep{astropy:2013, astropy:2018} and has made use of NASA’s Astrophysics Data System Bibliographic Services.

\section*{Data Availability}
The data underlying this article will be shared on reasonable request to the corresponding author.


\bibliographystyle{mnras}
\bibliography{bibliography}


\bsp	
\label{lastpage}
\end{document}